\documentclass[onecolumn,showpacs,preprintnumbers,amsmath,amssymb]{revtex4}
\usepackage{graphicx}
\usepackage{braket}

\begin{document}
\title{\bf Non-adiabatic Non-cyclic Generalization of the Berry Phase for a Spin-$1/2$ Particle in a Rotating Magnetic Field}
\author{S.S. Gousheh}
\email{ss-gousheh@sbu.ac.ir}
\author{A. Mohammadi}
\email{a_mohammadi@sbu.ac.ir}
\author{L. Shahkarami}
\email{l_shahkarami@sbu.ac.ir}
\affiliation{%
Department of Physics, Shahid Beheshti University G. C., Evin, Tehran
19839, Iran
}%
\date{\today}

\begin{abstract}
In this paper we define a non-dynamical phase for a spin-$1/2$ particle in a rotating magnetic field in the non-adiabatic non-cyclic case, and this phase can be considered as a generalized Berry phase. We show that this phase reduces to the geometric Berry phase, in the adiabatic limit, up to a factor independent of the parameters of the system. We could add an arbitrary phase to the eigenstates of the Hamiltonian due to the gauge freedom. Then, we fix this arbitrary phase by comparing our Berry phase in the adiabatic limit with the Berry's result for the same system. Also, in the extreme non-adiabatic limit our Berry phase vanishes, modulo $2\pi$, as expected. Although, our Berry phase is in general complex, it becomes real in the expected cases: the adiabatic limit, the extreme non-adiabatic limit, and the points at which the state of the system returns to its initial form, up to a phase factor. Therefore, this phase can be considered as a generalization of the Berry phase. Moreover, we investigate the relation between the value of the generalized Berry phase, the period of the states and the period of the Hamiltonian.
\end{abstract}
\maketitle

\section{Introduction}
Berry's profound discovery \cite{berry} of the geometrical phase has attracted a great amount of interest since 1980s
. He considered a quantum mechanical system whose Hamiltonian $H(\mathbf{R})$ depends on external parameters $\mathbf{R}(t)$, which are supposed to vary slowly in time in order for the adiabatic theorem to hold. The adiabatic theorem \cite{Messiah} guarantees that if the system is initially in an instantaneous eigenstate of the Hamiltonian, it continues to remain in the same instantaneous eigenstate of the
 Hamiltonian $H(\mathbf{R}(t))$, as time evolves. If $\mathbf{R}$ is transported around a circuit $C$ and returns to its initial state, obviously the Hamiltonian does the same, and by the adiabatic theorem so does the state of the system up to
a phase factor. This phase factor has a geometrical part, $\exp (i\gamma(C))$, in addition to the familiar dynamical part which comes from the factor $\exp (-iEt/\hbar)$.
 Berry has calculated $\gamma(C)$ for a general adiabatic case in which the change in the Hamiltonian is cyclic. Then, He used it for spinning particles in a slowly-changing magnetic field as a demonstration.
\par An important type of evolution of a physical system is a cyclic evolution which is conventionally defined as one in which the state of the system returns to its original state.
Aharonov and Anandan \cite{Aharonov} obtained a generalized Berry phase by investigating the phase change for these cyclic evolutions without the restriction of adiabaticity. They have obtained a geometrical phase which is universal in the sense that it is the same for the infinite number of
curves in the Hilbert space $\mathcal{H}$ which is projected into a unique curve $C$ in the projective Hilbert space $\mathcal{P}$. Unlike
the case studied by Berry, their initial state does not have to be an eigenstate of the Hamiltonian. Also, in their work the curve is not
necessarily closed in the parameter space, in order to have a cyclic evolution. The Berry results were retrieved in the adiabatic limit.
\par Simon \cite{Simon2} investigated the relation between the holonomy in a Hermitian line bundle and the Berry phase with the aid of the adiabatic theorem. As he has implied, since the Berry phase is the integral of the curvature, this phenomenon can be described by the holonomy. This realization gives a more compact formulation than that of Berry and the calculations are simpler in this formalism.
\par Samuel and Bhandari \cite{Samuel} generalized Berry's work to the non-unitary non-cyclic case. They have carried over Pancharatnam's ideas \cite{Pancharatnam} to quantum mechanics to get this extension of the Berry phase.
\par Wang \cite{Wang} employed a special model to investigate the non-adiabatic cyclic Berry phase, in detail. The system was a particle with
an arbitrary spin $j$ in a rotating magnetic field. He chose a special initial state so that it returns to its original state
when the Hamiltonian does so. With this choice, this model becomes an exactly solvable one. In this model also the Berry phase can
be obtained in the appropriate limit.
\par After all the works done for the geometric phase in pure states, this phase was extended to mixed states by Uhlmann \cite{Uhlmann}.
Since then, the mixed state geometrical phase, its generalization to non-unitary evolution and its experimental demonstrations have been considered by many other
works \cite{Sjoqvist,Singh,Tong,Bassi,Goto,Buric,Sjoqvist1}. Another issue of the geometric phase that has attracted interest, is the relation of the bipartite or multipartite
system with its subsystems. The discussions on the geometric phases of composite systems under local unitary evolutions \cite{Tong1,Tong2,Williamson,Xing,Li}
show that the geometric phase of the composite system does not always equal the sum of the geometric phase of its subsystems. Niu et al. \cite{Niu}, investigating the model of two interacting spin-half particles
in a rotating magnetic field, have given the relation between the geometric phase of the composite system and its subsystems in non-local unitary evolution.
\par The first experimental verification of Berry's geometrical phase has been performed by Tomita and Chiao \cite{Tomita}. They have proved the existence of the Berry phase by use of a single-mode, helically wound optical fiber, inside which a linearly polarized photon is transported adiabatically around a closed path in momentum space.
They carried out this experiment on the basis of Chiao and Wu's prediction \cite{Chiao}. This experiment, which is essentially at the classical level, has also shown "the topological nature" of this phase. After that, some other experiments have been carried out on this subject \cite{Simon,Bitter,hasegawa,filipp,sponar}. For a comprehensive review please see `The Geometric Phase in Quantum Systems' by Ali Mostafazadeh et. al. \cite{Mostafazadeh}.
\par In this paper we employ an exactly solvable model in order to fully investigate some of the profound principles of the Berry phase. The specific principles that we have in mind are the implications resulting from the alleviation of the adiabatic restriction. The model that we choose is the usual spin-$1/2$ particle in a rotating magnetic field. In section 2, we introduce the model and solve it exactly without the adiabatic restriction. We exhibit the particular solution when the initial condition is one of the two instantaneous eigenstates of the Hamiltonian. We then plot the probability of finding the system in the initial eigenstate at the period of the Hamiltonian. In section 3, we define a generalized Berry phase for this model. We show that this phase has the correct adiabatic and extreme non-adiabatic limits. We also plot the real and imaginary parts of our Berry phase as a function of time and find that the imaginary part is zero at the state periods. Furthermore, we find the specific conditions under which the Hamiltonian and the state of the system are simultaneously cyclic. Finally, we discuss the relation between this generalized phase, the period of the Hamiltonian and the period of the state.
\section{The Model}
\begin{center}
\begin{figure}[th] \includegraphics[width=7cm]{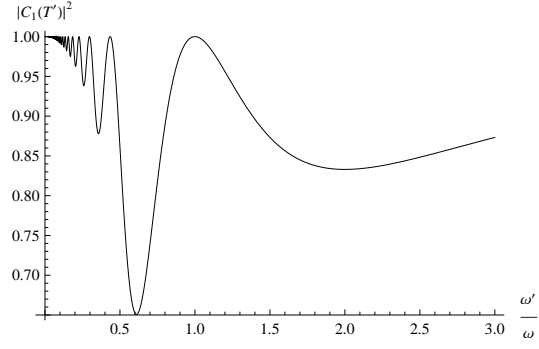}\caption{\label{c-omega} \small
  The probability of finding the system in the state $|1(T')\rangle $ as a function of $\omega'/\omega$ for $\cos\beta=1/2$.}
  \label{geometry}
\end{figure}
\end{center}

\begin{center}
\begin{figure}[th] \includegraphics[width=7cm]{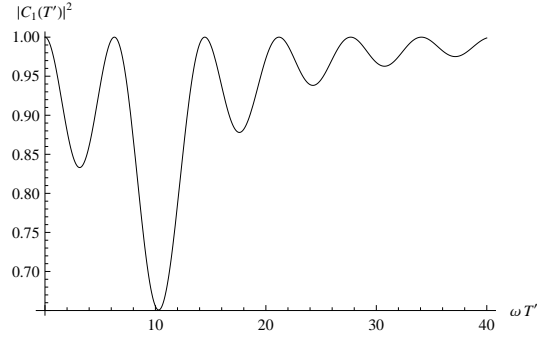}\caption{\label{c-tprime} \small
  The probability of finding the system in the state $|1(T')\rangle $ as a function of $\omega T'$ for $\cos\beta=1/2$.}
  \label{geometry}
\end{figure}
\end{center}

Consider a system with a Hamiltonian $H(t)$ possessing a set of instantaneous
eigenstates ${|n(t)\rangle}$, i.e. $H(t)|n(t)\rangle=E_n(t)|n(t)\rangle$. In general, even if the initial state of the system is one of its eigenstates namely ${|n(0)\rangle}$, one does not expect the system to remain in that instantaneous
eigenstate ${|n(t)\rangle}$ with the passage of time, except in the adiabatic limit. In general, the state of the system could be expanded in terms of the instantaneous eigenstates of the
Hamiltonian, that is,
\begin{equation}\label{e1}
|\psi(t)\rangle=\sum_n C_n(t)|n(t)\rangle.
\end{equation}
Substituting this state into the time-dependent Schr\"{o}dinger equation, we obtain,
\begin{equation}\label{e3}
E_n(t)C_n(t)-i\hbar \dot{C}_n(t)-\sum_m i\hbar C_m(t)\langle n(t)|\frac{dm(t)}{dt}\rangle=0.
\end{equation}
 The solution to this set of equations leads to the coefficients $C_n(t)$.
 As an example, consider a spin-$1/2$ system in a magnetic field which rotates about the $z$ axis with the angular frequency $\omega'$ and that is in an arbitrary direction
 $\hat{n}=(\sin\beta \cos\alpha,\sin\beta \sin\alpha,\cos\beta)$ at $t=0$. Then, at arbitrary time $t$ we obtain,
\begin{equation}\label{e5}
\vec{B}(t)=|\vec{B}|(\sin\beta \cos\alpha'(t),\sin\beta \sin\alpha'(t),\cos\beta),
\end{equation}
where $\alpha'(t)=\alpha+\omega't$. The Hamiltonian can be written as,
 \begin{equation}\label{e8}
H(t)=-\frac{e}{m_ec}\vec{S}.\vec{B}(t)=\frac{\hbar\omega}{2}\begin{pmatrix}\cos\beta & \sin\beta e^{-i\alpha'(t)} \\ \sin\beta e^{i\alpha'(t)} & -\cos\beta
\end{pmatrix},
\end{equation}
The instantaneous eigenstates and eigenvalues of the Hamiltonian $H(t)$ are as follows
\begin{equation}\label{e12}
\begin{split}
&|1(t)\rangle=\cos\frac{\beta}{2}\exp\left(\frac{-i}{2}\alpha'(t)-i\delta_1(t)\right)|+\rangle+
\sin\frac{\beta}{2}\exp\left(\frac{i}{2}\alpha'(t)-i\delta_1(t)\right)|-\rangle \quad,\quad E_1(t)=\frac{\hbar\omega}{2},\\
&|2(t)\rangle=\sin\frac{\beta}{2}\exp\left(\frac{-i}{2}\alpha'(t)-i\delta_2(t)\right)|+\rangle-
\cos\frac{\beta}{2}\exp\left(\frac{i}{2}\alpha'(t)-i\delta_2(t)\right)|-\rangle \quad,\quad E_2(t)=-\frac{\hbar\omega}{2}.
\end{split}
\end{equation}
The appearance of the arbitrary phases $\delta_1(t)$ and $\delta_2(t)$ are due to the freedom embedded in the instantaneous eigenvalue equation. From Eq.\,(\ref{e3}) for this system, we have
\begin{equation}\label{e10}
\begin{split}
&E_1(t)C_1(t)-i\hbar \dot{C}_1(t)- i\hbar C_1(t)\bigg \langle 1(t) \bigg| \frac{d1(t)}{dt}\bigg\rangle- i\hbar C_2(t)\bigg\langle 1(t)\bigg|\frac{d2(t)}{dt}\bigg\rangle=0,\\
&E_2(t)C_2(t)-i\hbar \dot{C}_2(t)- i\hbar C_1(t)\bigg\langle 2(t)\bigg|\frac{d1(t)}{dt}\bigg\rangle- i\hbar C_2(t)\bigg\langle 2(t)\bigg|\frac{d2(t)}{dt}\bigg\rangle=0.
 \end{split}
\end{equation}
Inserting Eqs.\,(\ref{e12}) into Eqs.\,(\ref{e10}), we obtain
\begin{equation}\label{e13}
\begin{pmatrix}\dot{C}_1(t)\\\dot{C}_2(t)
\end{pmatrix}=\begin{pmatrix}-i\frac{\omega}{2}+i\frac{\omega'}{2}\cos\beta+i\dot{\delta_1} & i\frac{\omega'}{2}\sin\beta \textrm{e}^{i(\delta_1-\delta_2)} \\
i\frac{\omega'}{2}\sin\beta \textrm{e}^{-i(\delta_1-\delta_2)} & i\frac{\omega}{2}-i\frac{\omega'}{2}\cos\beta+i\dot{\delta_2}
\end{pmatrix}\begin{pmatrix}C_1(t)\\C_2(t)
\end{pmatrix}.
\end{equation}
We choose $\delta_1(t)=\delta_2(t)=A+B\omega't$, since this at the most amounts to a constant shift in $H(t)$ and we fix the coefficients latter by making a correspondence with the adiabatic limit. Any higher powers of time, will invalidate the overall properties of the time evolution operator. Now we assume that the initial state of the system is $|1(0)\rangle$. Then, solving
Eqs.\,(\ref{e13}) with this condition, we obtain
\begin{equation}\label{e14}
\begin{split}
C_1(t)=&\frac{1}{\lambda}
\bigg\{\left (\frac{\omega}{2}-\frac{\omega'}{2}\cos\beta-\frac{\lambda}{2}\right)
\exp\left(\frac{it}{2}(2B\omega'-\lambda)\right) \\
&-\left (\frac{\omega}{2}-\frac{\omega'}{2}\cos\beta+\frac{1}{2}\lambda\right)
\exp\left(\frac{it}{2}(2B\omega'+\lambda)\right)\bigg\},\\
C_2(t)=&\frac{-\frac{\omega'}{2}\sin\beta}{\lambda}
\left\{\exp\left(\frac{it}{2}(2B\omega'-\lambda)\right)
-\exp\left(\frac{it}{2}(2B\omega'+\lambda)\right)\right\},
\end{split}
\end{equation}
where $\lambda=\sqrt{\omega^2+\omega'^2-2\omega\omega'\cos\beta}$. The state of the system at an arbitrary time $t$ is
\begin{equation}\label{e15}
|\psi(t)\rangle=C_1(t)|1(t)\rangle+C_2(t)|2(t)\rangle.
\end{equation}
Figure (\ref{c-omega}) shows the $|C_1(T')|^2$, the probability of
finding the system in the state $|1(t)\rangle$ at $t=T'$, as a function of $\omega'/\omega$, where $T'$ is the period of the Hamiltonian. As one expects, in the limit of the adiabatic
evolution of the Hamiltonian, the Berry limit, $|C_1(T')|^2$ tends to $1$. This means that for
$\omega'/\omega\rightarrow0$, the system is able to correlate itself with the Hamiltonian
instantly and remains in the state which it begins with. Also, in the extreme non-adiabatic limit the probability of finding the system in $|1(T')\rangle$ is $1$. This means that in the limit that the Hamiltonian changes extremely rapidly, the instantaneous state of the system is incapable of responding to the changes. As can be seen in Fig.\,(\ref{c-omega}), there are also other points at which $|C_1(T')|^2=1$. At these points the system's state returns to the initial state after one period of
the Hamiltonian. In Fig.\,(\ref{c-tprime}), $|C_1(T')|^2$ has been plotted as a function of $\omega T'$.
The intercept that is $\omega T'=0$ indicates the extreme non-adiabatic limit in which the state of the system does not change. The subsequent maxima indicate the points at which the state of the system has returned to its original state. At the first maximum, that is $\omega T'=2\pi$, the state returns to its initial state
after one cycle. In general, if we label these maxima by the set $m=\{0,1,2,3,...\}$, then $m$ indicates the number of times the state of the system has cycled. It is important to note that only for certain points (with measure zero) will the cycles of the Hamiltonian and the state match.

\section{The Non-adiabatic Non-cyclic Generalization of Berry Phase}
\begin{center}
\begin{figure}[th] \includegraphics[width=7cm]{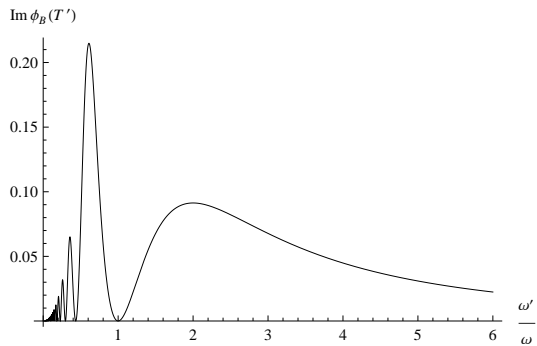}\caption{\label{imphib} \small
  The imaginary part of the Berry phase as a function of $\omega'/\omega$ for $\cos \beta=1/2$.}
  \label{geometry}
\end{figure}
\end{center}

\begin{center}
\begin{figure}[th] \includegraphics[width=7cm]{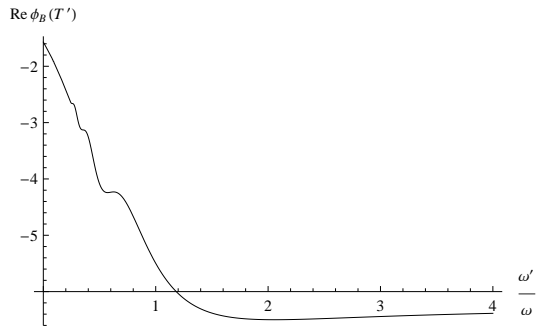}\caption{\label{rephibnim} \small
 The real part of the Berry phase as a function of $\omega'/\omega$ for $\cos \beta=1/2$.}
  \label{geometry}
\end{figure}
\end{center}

\begin{center}
\begin{figure}[th] \includegraphics[width=7cm]{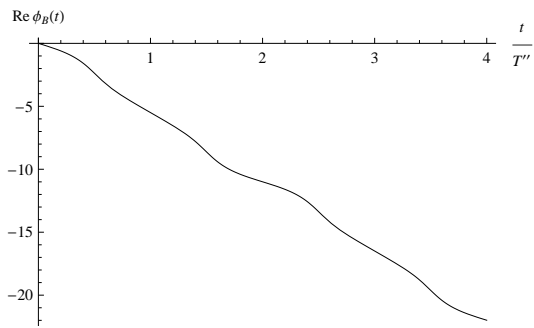}\caption{\label{rephibtstate} \small
  The real part of the Berry phase as a function of $t/T''$ for $\cos \beta=1/2$ at $\omega'=\omega$.}
  \label{geometry}
\end{figure}
\end{center}

\begin{center}
\begin{figure}[th] \includegraphics[width=7cm]{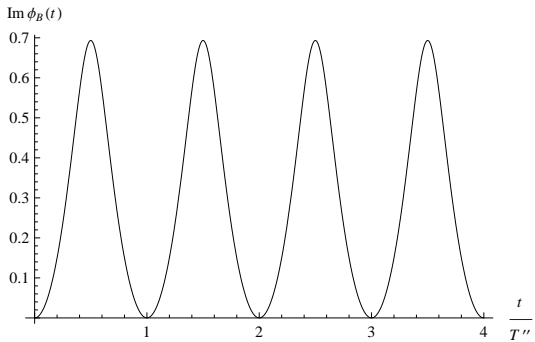}\caption{\label{Imphistate} \small
  The imaginary part of the Berry phase as a function of $t/T''$ for $\cos \beta=1/2$ at $\omega'=\omega$.}
  \label{geometry}
\end{figure}
\end{center}

\begin{center}
\begin{figure}[th] \includegraphics[width=7cm]{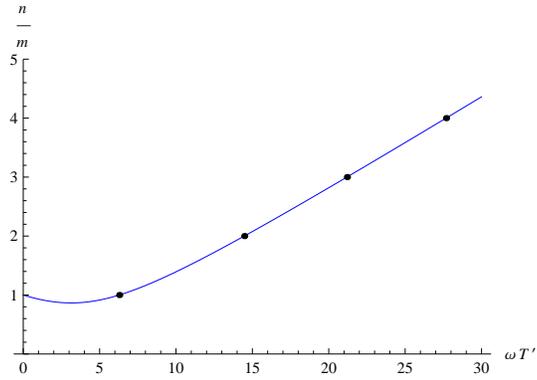}\caption{\label{nm} \small
 $n/m$ as a function of $\omega T'$ for $\cos \beta=1/2$.}
  \label{geometry}
\end{figure}
\end{center}

In the original paper of Berry, the Hamiltonian is assumed to change adiabatically in time. Then, at any instant, the system will remain in the same instantaneous eigenstate which it starts with, apart from a phase factor, in accordance with the adiabatic theorem. Subtraction of the dynamical phase from this phase factor leads to the geometrical Berry phase which is real. To define the Berry phase in the non-adiabatic non-cyclic case, we can set $C_1(t)=e^{i\theta}$ in which $\theta$ can in general be complex. Then, we subtract the dynamical phase from this phase factor to obtain the non-adiabatic non-cyclic Berry phase that might not be real anymore. However, as we shall show, this generalized Berry phase is real when the state is cyclic, and this includes both the adiabatic and the extreme non-adiabatic limits. Therefore, we have
\begin{equation}\label{e16}
C_1(t)=\exp(i\theta)=\exp(-\theta_i)\exp(i\theta_r),
\end{equation}
where
\begin{equation}\label{e17}
\theta_r=-i\ln\left(\frac{C_1(t)}{|C_1(t)|}\right),\quad\quad \theta_i=-\ln|C_1(t)|.
\end{equation}
A nonzero value for $\theta_i$ indicates that the probability of finding the system in the other states is nonzero. The dynamical phase is defined as follows
\begin{equation}\label{e18}
\phi_D(t)=-\frac{1}{\hbar}\int_0^{t}\langle\psi(t)|H(t')|\psi(t')\rangle dt'.
\end{equation}
We define the generalized Berry phase by
\begin{equation}\label{e2}
\theta(t)=\phi_D(t)+\phi_B(t).
\end{equation}
Then,
\begin{equation}\label{e19}
\begin{split}
\phi_B(t)&=\theta_r(t)+i\theta_i(t)-\phi_D(t)\\
&=B\omega't-i \ln \left\{\frac{\lambda \cos(\frac{\lambda}{2}t)-i\sin(\frac{\lambda}{2}t)(\omega-\omega'\cos\beta) }{\sqrt{\lambda^2 \cos^2(\frac{\lambda}{2}t)+\sin^2(\frac{\lambda}{2}t)(\omega-\omega'\cos\beta)^2}}\right\}\\
&-\frac{i}{2} \ln \left[\lambda^2 \cos^2(\frac{\lambda}{2}t)+\sin^2(\frac{\lambda}{2}t)(\omega-\omega'\cos\beta)^2\right ]\\
&+\frac{\omega}{2}t\left[1-\frac{\omega'^2}{\lambda^2}\sin^2 \beta\right ]
+\frac{\omega}{2\lambda}\sin (\lambda t)\left[\frac{\omega'^2}{\lambda^2}\sin^2 \beta\right ].
\end{split}
\end{equation}
From this point on we restrict our calculations to the case $t=T'$, that is the time elapsed equals one period of the Hamiltonian. Only the third term in the expression for the generalized Berry phase (Eq.\,(\ref{e19})) is imaginary. It can be easily shown that this term tends to zero when $\omega'\rightarrow 0,\infty$.
Furthermore, At the cyclic points where the state of the system returns to its initial state, we have $|C_1(T')|^2=1$ and $\mbox{Im}\, \varphi_B=0$. This means that our definition of the Berry phase for these cases is real, like the other established results. These results are shown in the Fig.\,(\ref{imphib}).
Our non-adiabatic Berry phase has to approach the Berry's result in the adiabatic limit ($\omega'\rightarrow0$), which is $\phi_B=\pi \cos\beta-\pi$ for this special example. Therefore, we should consider the real part of our Berry phase in the Eq.\,(\ref{e19}) at $t=T'$ and in the limit $\omega'\rightarrow0$. Comparing the Berry's result and ours, that is $\lim_{\omega'\rightarrow0}\text{Re}\,\phi_B (T')=2\pi B+\pi \cos \beta$, we obtain $B=-1/2$. Figure (\ref{rephibnim}) depicts the real part of the Berry phase for the case $\cos \beta=1/2$.
It can be easily shown that in the extreme non-adiabatic limit for the arbitrary $\beta$, $\mbox{Re}\, \phi_B\rightarrow-2\pi$, that is $\exp (i\theta)\rightarrow\exp (-2i\pi)=1$. This result is expected since in the extreme non-adiabatic limit, the state of the system cannot match itself with the changes.
In the adiabatic case considered by Berry the geometrical phase was calculated at the period of the Hamiltonian and the system remained in its initial state, up to a phase factor. However, here we consider the non-adiabatic non-cyclic case in which the system gets out of its initial state and it can go to any state in the Hilbert space. Therefore, it is interesting to look closely at the system and find its period. We obtain the state period $T''$ by setting $C_2(t)=0$, that is
\begin{equation}\label{e23}
C_2(T'')=\frac{i \omega' \sin\beta}{\lambda}
\\
\sin\left(\frac{T''}{2}\lambda\right)\textrm{e}^{-i\omega' T''/2}=0.
\end{equation}
Thus,
\begin{equation}\label{e24}
T''=\frac{2\pi}{\lambda}.
\end{equation}
Therefore, $C_1(nT'')$ is a pure phase factor, i.e. the phase is real. Thus, $\phi_B$ is real since the dynamical phase is always real. Figures (\ref{rephibtstate}) and (\ref{Imphistate}) show the real and imaginary parts of the Berry phase as a function of $t/T''$ for the case $\cos \beta=1/2$ at $\omega'=\omega$, respectively. As expected the imaginary part of the Berry phase is zero at the state period, that is the points at which $t/T''$ is integer. Also, it can be seen from Fig.\,(\ref{rephibtstate}), $\text{Re}\,\phi_B(t)=0$ at $t=0$, as expected.
We can now investigate the relation between the state's period and the Hamiltonian's period. Generally these are not equal, however if we insist that both the Hamiltonian and the state be cyclic, we must impose $nT''=mT'$, hence
\begin{equation}\label{e25}
\frac{2m\pi}{\sqrt{\omega^2+\omega'^2-2\omega\omega' \cos\beta}}=\frac{2n\pi}{\omega'},
\end{equation}
\begin{equation}\label{e26}
\frac{n}{m}=\frac{\sqrt{4 \pi^2+\omega^2T'^2-4\pi\omega T' \cos\beta}}{2\pi}.
\end{equation}
This relation represents when the Hamiltonian cycles $m$ times, the state cycles $n$ times so that both of them return to their initial forms at the same time.
In Fig.\,(\ref{nm}) we have plotted the Eq.\,(\ref{e26}). Let us consider the case $m=1$, that is the points in Fig.\,(\ref{nm}) in which $n/m$ is integer. As one expects, these points correspond to the maxima of $|C_1(T')|^2$ in Fig.\,(\ref{c-tprime}). In the same way, we can see that there is always some $T'$ which satisfies the Eq.\,(\ref{e26}) for any chosen integer value of $n$, $m$.

\section{Conclusion}
 In this paper, we have analyzed the time evolution of a system of a spin-$1/2$ particle in a rotating magnetic field in full details. We have defined a non-adiabatic non-cyclic generalization of the Berry phase. Since the model is exactly solvable, we have been able to analyze completely this phase. We have added an arbitrary phase to the eigenstates of the Hamiltonian and then fixed it by comparing our Berry phase in the adiabatic limit with the Berry's result for the same system. This result gives the correct result in the extreme non-adiabatic limit. Our Berry phase is in general complex. However, it becomes real in the expected cases: the adiabatic limit, the limit $\omega'\rightarrow\infty$, and the points at which the state of the system returns to its initial form, up to a phase factor. We have also plotted the real and imaginary parts of our Berry phase as a function of time and find that the real part is zero at the initial time, as expected. Furthermore, we have derived a relation which holds when both the Hamiltonian and the state are cyclic simultaneously.

 \end{document}